# The Influence of Teamwork Quality on Software Team Performance

Emily Weimar,[1,3] Ariadi Nugroho,[3] Joost Visser,[2,3] Aske Plaat,[1] Martijn Goudbeek,[1] Alexander P. Schouten[1]

[1] Tilburg University;  [2] Radboud University, Nijmegen;  [3] Software Improvement Group, Amsterdam

# The Influence of Teamwork Quality on Software Team Performance

ABSTRACT

Traditionally, software quality is thought to depend on sound software engineering and development methodologies such as structured programming and agile development. However, high quality software depends just as much on high quality collaboration within the team. Since the success rate of software development projects is low (Wateridge, 1995; The Standish Group, 2009), it is important to understand which characteristics of interactions within software development teams significantly influence performance. Hoegl and Gemuenden (2001) reported empirical evidence for the relation between teamwork quality and software quality, using a six-factor teamwork quality (TWQ) model. This article extends the work of Hoegl and Gemuenden (2001) with the aim of finding additional factors that may influence software team performance. We introduce three new TWQ factors: trust, value sharing, and coordination of expertise. The relationship between TWQ and team performance and the improvement of the model are tested using data from 252 team members and stakeholders. Results show that teamwork quality is significantly related to team performance, as rated by both team members and stakeholders: TWQ explains 81% of the variance of team performance as rated by team members and 61% as rated by stakeholders. This study shows that trust, shared values, and coordination of expertise are important factors for team leaders to consider in order to achieve high quality software team work.

## 1. Introduction

Traditionally, software quality is thought to depend on sound software engineering practices and development methodologies (such as structured programming or agile development). However, software quality has also been shown to depend on good teamwork, specifically with respect to the interaction processes within a team (Hoegl & Gemuenden, 2001; Liang, Wu, Jiang & Klein, 2012; Henderson & Lee, 1992). Extensive research, see, for example, Wateridge (1995) and The Standish Group (2009), has shown a low success rate of 32% of information system development projects; 44% of the projects surpassed the planned budget and time; and 24% of the projects failed completely. Since software development is primarily a team effort (Faraj & Sproull, 2000), it is important to understand the factors or characteristics in software development teams that influence team performance.

High quality teamwork is considered a crucial success factor in software development projects (Cooper, 1993; Pinto, Pinto & Prescott, 1993; Gemuenden, 1990; Griffin & Hauser, 1992). However, few studies have empirically investigated the exact characteristics of teamwork that influence the success of innovative teams However, in 2001 Hoegl and Gemuenden (2001) studied these interactions, presenting the first empirical indication of the impact of team work quality on software quality. They present two main points of critique of earlier studies: (1) previous research did not address the multifaceted nature of teams but rather focused on the relationship between team-based organizations and performance (Gupta, Ray, & Wileman, 1987; Hise, O'Neal, Parasuraman, & McNeal, 1990; Cooper & Kleinschmidt, 1995); and (2) there are conflicts in the literature about the impact of teamwork on team success (Thamhain & Kamm, 1993; Campion, Medsker & Higgs, 1993; Cohen, Ledford & Spreitzer, 1996). In order to address these issues, Hoegl and Gemuenden (2001) studied the influence of six teamwork quality (TWQ) factors – *viz.* communication, coordination, balance of member contribution, mutual support, effort, and cohesion – on the success of innovative projects. They based their model on the fundamental idea that the success of teams depends on the degree to which team members are able to collaborate. The results were promising; the TWQ factors significantly correlated with performance ratings, explaining 41% of the variance of team member ratings, 11% of team leader ratings, and 7% of the manager ratings.

In discussing what other factors might explain the variance, Hoegl and Gemuenden (2001) argue that teamwork quality is not the only determinant for project success, and suggest other factors such as management, organizational factors and communication between the team and external sources might also play a role in determining project success. In this study we propose three different additional factors that may further explain high quality team performance. Trust, for example, is found to be a key predictor for team performance (Friedlander, 1970) and an important support mechanism for teamwork (Bandow, 2001; Salas, Sims & Burke, 2005), but it is not part of Hoegl and Gemuenden's (2001) model. Also, Hoegl and Gemuenden (2001) measured team uniformity related to work norms concerning effort. We argue that effort is one of the multiple facets of work norms. Therefore, we propose to use a more generic measure of work norms, including uniformity of values, goals and beliefs. Hence, covering norms about effort.

Our proposed new model includes the following teamwork quality factors: communication, coordination of expertise, cohesion, trust, mutual support, and shared values (see Figure 2; the original model is depicted in Figure 1). The goal of the research, therefore, is to validate if the following assumption holds: *(1) Trust, (2) Value Sharing, and (3) Coordination of Expertise are relevant factors for inclusion in Hoegl and Gemuenden's model of Teamwork Quality (TWQ).*

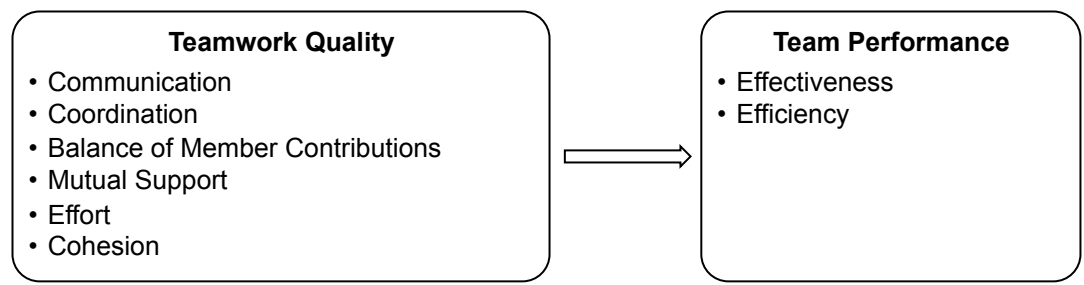

Figure 1. *Original TWQ model of Hoegl & Gemuenden (2001)*

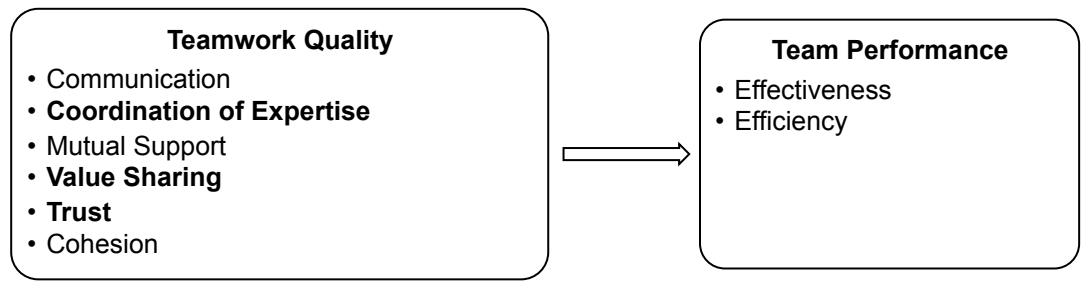

Figure 2. *New TWQ model (2013)*

We investigated the new teamwork quality model with 29 software development teams from 18 companies and compare the results with the performance of the teams in achieving their goals. All the team members that have a direct contribution to software development, from the analysis to the testing phase, were asked to participate in the study. Project performance was measured by collecting data about the success of the team on different measures (i.e.,, quality, schedule, costs, and morale) from both team members and project stakeholders. We then (1) statistically analyzed the relationship between team characteristics and team performance, (2) compared the explanatory power of the new model with that of Hoegl and Gemuenden (2001), and (3) tried to explain the differences between the models. This article provides additional support for the notion that better teamwork creates better software. The results of our study support three main findings: (1) There is a significant relationship between teamwork quality and the success of software development projects as measured by team performance (effectiveness and efficiency); (2) The magnitude of the relationship between TWQ and team performance differs with the perspective of the rater (team member versus stakeholder): the new TWQ explains 81% of the variance of team performance as rated by team members and 61% as rated by stakeholders; (3) The addition of trust, coordination of expertise, and value sharing to the TWQ model of Hoegl and Gemuenden (2001) has increased the explanatory power. When contrasted to the original TWQ model, the new TWQ appears to do a better job at explaining the available variance in ratings of team performance by team members (66% versus 40%) as well as by stakeholders (40% versus 11%).

## 2. THEORETICAL FRAMEWORK

In this section, we will first elaborate on the definitions of teams, teamwork and performance. Second, we will discuss the TWQ factors (Table 1) and their relationship or contribution to the complete TWQ model, based on current scientific literature on teamwork, software development and team performance.

### 2.1 TEAMS, TEAMWORK AND PERFORMANCE

Table 1

*Teamwork quality factors*

| |
| --- |
| **Communication**. There is sufficient, frequent, spontaneous, timeliness, precise and useful exchange of information. |
| **Coordination of Expertise**. Location and need of expertise are known and coordinated. |
| **Cohesion**. Team members are motivated to maintain the team and there is team spirit. |
| **Trust**. Team members trust each other. |
| **Mutual Support**. Team members help and support each other in carrying out their tasks. |
| **Value Sharing**. Team members share the same values and goals. |
| **Team Performance**. The degree to which the project team completes the project efficiently and effectively. |

Team performance can be assessed in terms of effectiveness and efficiency. *Effectiveness* is the degree to which a team meets the expectations of the quality of the outcome (Hackman, 1987). This refers to, for example, the degree to which the goals and quality of the project were met. *Efficiency* refers to the degree to which the team met time and budget objectives (Hoegl & Gemuenden, 2001). Effectiveness, therefore, concerns the comparison between actual and intended *outcome*, and efficiency refers to the assessment of the actual inputs compared to the intended *inputs*.

### 2.2 SELECTION OF FACTORS

Hoegl and Gemuenden's TWQ model is one of the few empirically tested comprehensive models on team interaction and software quality. It is a frequently cited in the literature. In discussing their results, Hoegl and Gemuenden suggested including other relevant factors in the TWQ model. We agree. Based on existing social interaction literature, we have chosen to include trust, shared value and coordination of expertise in our model. We will now briefly discuss our reasons ,(Author-reference 2013) contains a more elaborate discussion. First, Hoegl and Gemuenden exclude trust. However, Friedlander (1970), and many others, argues that trust is a key predictor for team performance. Second, Hoegl and Gemuenden argue that especially norms about the effort of team members are important for TWQ. However, norms about effort only partly cover team norms and values; other norms and values such as those pertaining about manners, are important, too. Therefore, we propose to use a more encompassing factor of value sharing that covers both norms about the effort as well as other team norms and values. Third, Hoegl and Gemuenden's model contains *balance of member contributions*, which measures whether contributions to the team are balanced in terms of member's specific knowledge and experience. Faraj and Sproull's measure of coordination of expertise incorporates is again more encompassing, it measures (1) whether and where expertise is located within the team, (2) if the team recognizes any need for expertise, and (3) if expertise within the team is brought to bear. This measure covers both the factor of coordination and balance of member contributions.

The next subsection discusses how each factor relates to team performance and why we chose to incorporate each factor in our model.

## 2.3 TEAMWORK FACTORS

### 2.3.1 COMMUNICATION

It is widely recognized that communication (in the sense of information sharing) is a fundamental component of teamwork and project success (Katz & Allen, 1998; Griffin & Hauzer, 1992). Communication provides a means to exchange information, share ideas among team members, coordinate efforts and provide feedback (Pinto & Pinto, 1990). Lu, Xiang, Wang, and Xiaopeng (2010), for example, found that a lack of information sharing and the existence of misunderstanding between team members and stakeholders of a project are two main causes of project failure. Communication can be seen as a primary tool that is needed to create a high-performing team. Therefore, communication in the sense of information sharing should be part of the TWQ model.

### 2.3.2 COORDINATION OF EXPERTISE

Hoegl and Gemuenden (2001) argue that coordination is an important aspect of teamwork. It refers to the development of and agreement on task-related goals, with well-defined subgoals for each member. Expertise is an elementary resource in software development, which is not considered in the study of Hoegl and Gemuenden (2001). Coordination of expertise, a much broader concept, refers to the "management of knowledge and skill dependencies" (Faraj & Sproull, 2000, p. 1555). This includes knowing where expertise is situated within a team, recognizing the need for expertise, and bringing expertise to good use. Faraj and Sproull (2000) found a significant positive relationship between coordination of expertise and team performance. Their measure of coordination of expertise, is, to a certain degree, a combination of Hoegl and Gemuenden's (2001) measures of coordination and balance of member contributions, supplemented with the link to expertise. Since software development is knowledge work, we consider the link between coordination and expertise to be important. Therefore, we decided to use Faraj and Sproull's (2000) measure of coordination of expertise instead of Hoegl and Gemuenden's (2001) measures of coordination and balance of member contributions.

### 2.3.3 COHESION

Cohesion is defined as "an individual's sense of belonging to a particular group and his or her feeling of morale associated with membership in a group" (Bollen & Hoyle, 1990, p. 482). Without a sense of belonging, team members would not like to associate with the rest of their team, and without a sense of morale, they would not be motivated to achieve organizational goals and objectives. Three forces of cohesion were distinguished by Mullen and Copper (1994), namely: (1) interpersonal attraction of team members, (2) commitment to the team task, and (3) group pride-team spirit. We consider and expect cohesion to be an important teamwork quality factor for software development teams. An adequate feeling of togetherness and belonging is needed to achieve high quality collaboration (Mullen and Copper, 1994).

## 2.3.4 TRUST

Friedlander (1970) found that trust is a key predictor for team performance. Following Mayer et al. (1995), we define trust as "the willingness of a party to be vulnerable to the actions of another party based on the expectation that the other will perform a particular action important to the trustor, irrespective of the ability to monitor or control the other party." Trust is an important supporting mechanism of teamwork. It influences many team processes such as the willingness to share information, give substantial feedback and manage time correctly (Bandow, 2001). Team members will communicate more openly and will share information more freely when they trust their team members or feel trusted by others. When they have the feeling their contribution is not appreciated, the chance is higher that they will not share information (Boss, 1978; Zand, 1972; Jones & George, 1998; Bandow, 2001). Furthermore, trust has a positive effect on job satisfaction (Driscoll, 1978; Muchinsky, 1977), satisfaction of communication and the perceived accuracy of information given (Roberts & O'Reily, 1974), and on satisfaction of working with the group (Ward, 1997). Given the importance of trust as a supporting mechanism for teamwork, this factor cannot be eliminated, and we decided to include trust in our TWQ model.

## 2.3.5 MUTUAL SUPPORT

According to Tjosvold (1995), mutual support is an essential element of TWQ. Teamwork hinges on the idea of mutual support of the team members rather than the competition between them (Hoegl & Gemuenden, 2001). Competition between people can exert a positive influence on the motivation and performance of individual tasks. For interdependent tasks such as software development, however, cooperation or mutual support amongst team members is more important. Both quality and acceptance of ideas generated by members of the team increase when members cooperate (Cooke & Szumal, 1994). Mutual support, therefore, is an important element of teamwork and needed to be able to reach team goals. The better team members support each other, the more effective and efficient these goals can be reached.

2.3.6 VALUE SHARING

Different types of team diversity exist (e.g., informational, social category and value diversity). With each type of diversity, different types of challenges and opportunities are involved. Each has a different influence on team performance (Jehn et al., 1999). Value diversity arises when team members have a different perspective on the team's task, goal, or mission. Such differences can lead to relationship, task, or process conflicts (Jehn, 1994; Jehn et al., 1999). In addition to influencing teamwork, value diversity also influences team performance. Low value diversity is needed to be efficient, effective and sustain a high moral within the team (Jehn et al., 1999). Hoegl and Gemuenden (2001) argue that especially norms about the effort of team members are important for TWQ. However, effort is only one of the multiple facets team members might have shared expectations about. Value diversity also relates to the team goal and mission. When team members share the same mission or vision, it is likely that they will prioritize the task of the team and have the same ideas regarding work norms. Therefore, we propose to include both effort, shared mission, and shared vision in the concept of value sharing.

2.3.7 PROJECT PERFORMANCE

Assessing project performance is notoriously difficult because perceptions of project performance differ due to the different interests of the parties involved. Members of software development teams, for example, usually relate project success to the completion of the scope of the project, while external stakeholders usually use measures of target time and costs to assess project performance (Agarwal & Rathod, 2006). Evaluations of project performance, therefore, can vary across team members, team leaders and stakeholders, making it hard to measure team performance objectively. Even though the subjectivity of performance measures remains a disputed topic, multiple measures of performance can be found in the literature. Hoegl and Gemuenden (2001) emphasize the importance of multiple ratings of team performance, coming from sources both internal and external to the team to improve objectivity.

It is important to pre-determine project requirements (quality, time and costs) and assessment criteria to reduce ambiguity and subjectivity about the definition of success. Furthermore, it is important that these goals and assessment criteria are communicated well among the different parties that are involved (Wateridge, 1995). Project performance, consequently, is measured by the assessment of project success by the different parties involved.

## 2.4 RESEARCH QUESTION

Based on the conceptual model and the review of the literature, our research question is whether: *(1) Trust, (2) Value Sharing, and (3) Coordination of Expertise are relevant factors for inclusion in Hoegl and Gemuenden's model of Teamwork Quality (TWQ).*

## 3. STUDY DESIGN

In this empirical study, the relationship between TWQ and team performance as rated by team members and stakeholders was investigated. Both qualitative and quantitative data were gathered by means of a fully standardized online questionnaire.

Two groups of participants were involved: team members and stakeholders. Stakeholders are not formal members of the team but are directly affected by the output of the team. Stakeholders include, for example, project sponsors or managers who were responsible for the production and implementation of the information system. First we made sure that all respondents corresponding to one team were referring to the same set of individuals as a team by asking participants to identify the team being evaluated. Team members then filled in characteristics of themselves, the team and the project. Then, questions measured the quality of teamwork and team performance. Finally, two open-ended questions allowed participants to give comments or remarks. Stakeholders were only asked to identify the team, assess its team performance, and were allowed to give additional remarks at the end. The complete questionnaire for team member can be found in the supporting material and in (Author-reference, 2013).

### 3.1.1 DATA COLLECTION

Teams had to fulfill the following conditions: (1) they had to be a software development team (2) of at least three members (3) that is embedded in an organization (4) and whose members considered themselves to be a team. All development teams belonged to the domain of business software products.

Participation recruitment was done through snowball sampling, making use of our network to make contact with organizations to ask for their willingness to participate. The two main channels we used were the Dutch CIO Platform[1] and the network of the EQuA Project.[2] The CIO platform is the professional organization of Chief Information Officers in the Netherlands. The EQuA Project is a software quality management improvement project in which universities and IT companies cooperate. Through the CIO Platform, we recruited development teams from the public sector. Private sector development teams were reached via the EQuA project. Furthermore, we made use of the business contacts of a software company to send out invitations for our study. All of these initial participation requests were done through email.

### 3.1.2 PROCEDURES

Prior to sending out the questionnaires to all team members and stakeholders, personal meetings with all team managers/leaders were arranged. During these meetings, the background, objectives and procedures of the study were explained. By asking the team managers to distribute the questionnaire, we expected that team members felt a higher commitment and motivation to fill out the questionnaire. Participation, however, was voluntary.

Filling out the team member questionnaire took each team member about 10 to 15 minutes, while stakeholders spent approximately 5 minutes.

---

[1] http://www.cio-platform.nl
[2] http://www.equaproject.nl

### 3.1.3 PARTICIPANTS

The study included two groups of respondents: (1) team members and (2) stakeholders. In total, 352 questionnaires were filled out; 204 team member questionnaires and 128 stakeholder questionnaires. Participants who quit halfway or did not complete the questionnaire were excluded from the analyses. Participants who (sporadically) did not answer a question(s), but did complete the questionnaire were not excluded from the sample. In total 199 team member responses and 53 stakeholder responses were used in the analyses.

In total 252 valid participants were included in the study for analyses. These participants were divided over 29 teams from 18 Dutch organizations, representing different industries including business and finance, IT/ICT, high technology, and governmental/public institutions. Characteristics of the sample population can be found in table 2. Appendix A (Table 8) gives an overview of team characteristics.

Participants included software team members (N = 199), of which 91.0% are male (N = 181) and 9.0% are female (N = 18), and project stakeholders (N = 53), all male. The average age of the team members was 37.37 years ($SD$ = 9.48).

### 3.1.4 MEASURES

All respondents were asked to evaluate teamwork factors and performance of the team as a whole. After a careful development phase and a pre-test and adjustments improving the clarity of the items, the questionnaire was sent out.

We used existing, validated scales from the literature to measure the factors (table 3). All items were measured on a 5-point Likert scale ranging from 1 (strongly disagree) to 5 (strongly agree).

**Communication.** Communication was measured by 5 items, adopted from Liang, Wu, Jian & Klein (2012), based on one of the teamwork quality facets of Hoegl and Gemuenden (2001).

Table 2

*Characteristics of sample population*

| Characteristic | Category | N | Percentage |
|---|---|---|---|
| Gender | Male | 181 | 91.0 |
| | Female | 18 | 9.0 |
| Age | <30 | 57 | 22.6 |
| | 31-40 | 67 | 34.2 |
| | 41-50 | 53 | 27.0 |
| | >50 | 19 | 9.7 |
| Educational background | IT | 138 | 69.3 |
| | Non-IT | 49 | 24.6 |
| | Other | 12 | 6.0 |
| Experience in software engineering | <5 years | 45 | 22.6 |
| | 6-10 years | 50 | 25.1 |
| | 11-15 years | 41 | 20.6 |
| | >15 years | 62 | 31.2 |
| Core team role | Requirement analyst | 22 | 11.1 |
| | Architect | 12 | 6.0 |
| | Developer | 106 | 53.3 |
| | Tester | 22 | 11.1 |
| | Project Manager | 15 | 7.5 |
| | Other | 22 | 11.1 |
| Time in team | <6 months | 33 | 16.6 |
| | 6-12 months | 35 | 17.6 |
| | 12-18 months | 33 | 16.6 |
| | >18 months | 98 | 49.2 |
| Average team size | <5 months | 9 | 31.0 |
| | 5-10 months | 14 | 48.3 |
| | >10 months | 6 | 20.7 |
| Project duration | <5 months | 10 | 34.5 |
| | 5-10 months | 8 | 27.6 |
| | >10 months | 11 | 37.9 |
| Main team assignment | System new development | 17 | 58.6 |
| | System renovation | 5 | 17.2 |
| | System re-development | 1 | 3.4 |
| | System maintenance | 5 | 17.2 |
| | Other | 1 | 3.4 |
| Type of application | Business application | 22 | 75.9 |
| | Real-time application | 2 | 6.9 |
| | Other | 5 | 17.2 |
| Development method | Waterfall | 9 | 31.0 |
| | Incremental | 4 | 13.8 |
| | RUP | 0 | 0.0 |
| | Agile | 16 | 55.2 |

Table 3

*Sources of teamwork quality measures*

| Teamwork Factors | Items | Adapted from |
| --- | --- | --- |
| Communication | 5 | Lian, Wu, Jian & Klein (2012) |
| Coordination of Expertise | 11 | Faraj & Sproull (2000) |
| Cohesion | 6 | Chin, Salisbury, Pearson & Stollak (1999) |
| Trust | 5 | Jarvenpaa, Knoll & Leidner (2006) |
| Cooperation | 6 | Hoegl & Gemuenden (2001) |
| Value Sharing | 6 | Jehn (1994) |
| Project Performance | 7 | Jones & Harrison (1996) |

Questions included being focused on the frequency of the communication, its spontaneity, team members' satisfaction of the timeliness of the information they received, the precision of communication and the usefulness. All items were measured on a 5-point Likert scale ranging from 1 (strongly disagree) to 5 (strongly agree).

**Coordination of Expertise.** Coordination of expertise was measured by 4 items for knowing expertise location, 3 items for recognizing the need of expertise, and 4 items for bringing expertise to good use (Faraj & Sproull, 2000). **Cohesion.** Cohesion was measured with the Cohesion Measurement Scale (CMS) by Chin, Salisburry, Pearson, and Stollak (1999), adopted from the CMS of Bollen and Hoyle (1990). Participants were asked if they felt they belonged to the group, were happy to be part of the group, saw themselves as part of the group, considered the group to be one of the best around, felt they were member of the group, and if they were content to be part of the group. **Trust.** The degree of trust within a team was measured by 5 items, adopted from Jarvenpaa and Leidner (2006). Team members were asked if the people in their team were trustworthy, if they considered one another's feelings on the team, if team members were friendly, reliable and trustworthy. **Mutual Support.** Mutual support considered the degree to which team members supported each other, suggestions and contributions of other team members were respected and further developed, and the team was able to reach consensus regarding important issues. Coordination was measured with the 6 items of mutual support of Hoegl and Gemuenden (2001). **Value sharing.** Value sharing was measured by 6 items, adopted from Jehn (1994). Participants were asked if the values of all team members were similar, whether the team as a whole had similar work values, if the team as a whole had similar goals, if the team members had strongly held beliefs about what is important within the team, whether the team members had similar goals, and whether all member agreed on what was important to the team. High scores indicated low value diversity.

**Team performance.** Project performance was considered as the degree to which the project goals were achieved, the expected amount of work was completed, a high level of quality was delivered, the schedule was adhered, the operations were carried out efficiently and within time limits, and to which the budget was adhered (Jones & Harrison, 1996). These items, based on the scales developed by Henderson & Lee (1992), have been used in multiple studies to measure team performance (i.e., Wang, Chang, Jiang, & Klein, 2012; Liu, Chen, Chen, & Sheu, 2011).

To control for internal consistency of the scales, we executed a Cronbach alpha test. All the scales had high reliabilities, all Cronbach's alpha $\geq .79$.

### 3.2 DATA ANALYSIS

To justify data aggregation to the team level, we examined inter-rater agreement. We used the two-way random method, looking for the absolute agreement among raters. Overall, there was a strong inter-rater agreement within the teams. Given this high homogeneity of within-team ratings, we aggregated the data at the team level by calculating the arithmetic mean (see table 4). Furthermore, Factor analyses assessed whether all six factors belong to the same latent construct. Subsequently, we examined the correlations between the different variables.

Following Hoegl and Gemuenden (2001), the TWQ model was tested using structural equation modeling (SEM). [3] The database included measures from the 199 team member respondents, supplemented with columns containing the (aggregated) measures of team performance as rated by the stakeholders of the corresponding team. Therefore, SEM at the individual level was done using N = 199.

---

[3] SEM is a large sample technique wherefore sample size is preferably not less than 200 (Lei & Wu, 2007). Because of the small sample size at the team level, we conducted a SEM using measures at the individual level. Missing data were replaced with the arithmetic mean of the team. To overcome the database design problem of the unequal amount of team members and shareholders, causing a lot of missing data, we used team member's measures of TWQ and performance at the individual level and the (aggregated) team level measures of performance of the stakeholders. Similar to the procedure described in Hoegl and Gemuenden (2001), the model was also tested using the aggregated team level ratings (N = 29).

## 4. Results

### 4.1 Teamwork Quality as a Latent Construct

To evaluate if all six TWQ factors relate to the same latent construct, factor analyses were conducted. A principal component analysis (PCA) with no rotation was conducted on the team level using aggregated team member responses. The Kaiser-Meyer-Olkin (KMO) measure verified the sampling adequacy for the analysis, KMO = .82 ('meritorious' according to (Kaiser, 1974)), and all KMO adequacy for individual items were > .77, which is well above the acceptable limit of .5 (Field, 2009). Bartlett's test of sphericity ($\chi^2$ (15) = 159.16, $p$ < .001), indicates that correlations between items were sufficiently large for PCA. An initial analysis was run to obtain eigenvalues for each component in the data. Only the first component had an eigenvalue over Kaiser's criterion of 1 and explained 79.81% of the variance. The scree plot was very clear and justified retaining only one component for further analyses. This latent construct represents TWQ

Table 4

*Means, standard deviations, and reliabilities of the variables at the team level*

|  | Mean | SD | Alpha[b] | ICC[c] |
|---|---|---|---|---|
| TWQ[a] |  |  |  |  |
| Communication | 3.78 | 0.62 | .84 | .79 |
| Coo. of Exp. | 3.78 | 0.53 | .86 | .82 |
| Cohesion | 4.04 | 0.64 | .92 | .88 |
| Trust | 4.07 | 0.52 | .79 | .76 |
| Mutual Support | 3.99 | 0.56 | .88 | .85 |
| Value Diversity | 3.48 | 0.67 | .84 | .86 |
|  |  |  |  |  |
| Team Performance (Team Member Rating) |  |  |  |  |
| Effectiveness | 3.52 | 0.75 | .82 | .81 |
| Efficiency | 3.46 | 0.65 | .79 | .78 |
|  |  |  |  |  |
| Team Performance (Stakeholder Rating) |  |  |  |  |
| Effectiveness | 3.70 | 0.81 | .87 | - |
| Efficiency | 3.51 | 0.71 | .76 | - |

[a] = team member ratings
[b] = Cronbach's alpha coefficient
[c] = intra-class correlation coefficient

Table 5

*Teamwork quality as a latent construct: Factor analysis, regression analysis, and reliability analysis at the team level*

| TWQ | Factor Loading | Std. Regr. Coefficients | *P*-Value |
|---|---|---|---|
| Communication | .88 | .20 | 0.00 |
| Coordination of Expertise | .89 | .19 | 0.00 |
| Cohesion | .92 | .18 | 0.00 |
| Trust | .88 | .17 | 0.00 |
| Mutual Support | .92 | .16 | 0.00 |
| Value Diversity | .88 | .22 | 0.00 |
| Eigenvalue | 4.79 | | $R^2$ 100% |
| Variance explained (Factor TWQ) | 79.81% | Cronbach's alpha coefficient 0.95 | |

*Note.* N = 29

Following Hoegl and Gemuenden (2001), we executed further factor analysis at the individual level (N = 199). Analyses at the individual level were done to ensure that the results at the team level (N = 29) were not the result of inflated correlations due to the aggregation of the data. Hoegl and Gemuenden (2001) followed two procedures to test the TWQ factor structure at the individual level. In our study, we only execute the first validation of reliability test procedure of Hoegl and Gemuenden (2001) to test the TWQ factor structure at the individual level. A random sample of 29 responses was used for factor analyses. This process was repeated 15 times. All factor analyses at the individual level support the findings at the team level; with all KMO measures being ≥ .77, all eigenvalues being greater than the Kaiser's criterion of 1 and total variance explained ranging from 58.78% to 83.12%. The results of these analyses are in supporting material. Table 5 shows the factor loadings of the factor analysis at the team level and the standardized regression coefficients of a linear regression with the dependent variable being the (aggregated) TWQ construct. Table 6 presents the correlations between all TWQ factors and performance variables. All variables are significantly positively correlated to each other. As expected, all TWQ factors correlate highly to each other since they all refer to the same latent construct. Correlation plots of all TWQ factors with team performance as rated by stakeholders are shown in supporting material.

## 4.3 Teamwork Quality and Team Performance

We investigated two structural equation models using SPSS with AMOS 18. Appendix B (Figure 4) contains a graphical depiction of the model. In contrast to Hoegl and Gemuenden, we conducted our structural equation model analyses at the individual level (N = 199) because of sample size limitations at the team level (at which Hoegl and Gemuenden did their analysis). To investigate possible differences between the team level and the individual level, we compared the analysis presented here with the same analysis done at the team level (N = 29). This did not show any substantial differences between the two analyses. Here, we only present the results of the analysis at the individual level (N = 199).

Table 6

*Pearson correlations between the TWQ factors and team performance measures (N = 199)*

|  | 1 | 2 | 3 | 4 | 5 | 6 | 7 | 8 | 9 | 10 |
|---|---|---|---|---|---|---|---|---|---|---|
| TWQ[a] |  |  |  |  |  |  |  |  |  |  |
| (1) Communication | - |  |  |  |  |  |  |  |  |  |
| (2) Coo. of Exp. | .79** | - |  |  |  |  |  |  |  |  |
| (3) Cohesion | .78** | .77** | - |  |  |  |  |  |  |  |
| (4) Trust | .62** | .78** | .81** | - |  |  |  |  |  |  |
| (5) Mutual Support | .81** | .70** | .83** | .75** | - |  |  |  |  |  |
| (6) Value Diversity | .70** | .75** | .73** | .75** | .81** | - |  |  |  |  |
| Team Performance (Team Member Rating) |  |  |  |  |  |  |  |  |  |  |
| (7) Effectiveness | .83** | .74** | 63* | .60* | .65** | .62** | - |  |  |  |
| (8) Efficiency | .71** | .63** | .75** | .65** | .78** | .74** | .79** | - |  |  |
| Team Performance (Stakeholder Rating) |  |  |  |  |  |  |  |  |  |  |
| (9) Effectiveness | .66** | .52** | .62** | .57** | .77** | .66** | .58** | .60** | - |  |
| (10) Efficiency | .78** | .46* | .55** | .48** | .65** | .43* | .53** | .47* | .71** | - |

*Note.* $* p < .05$, $** p < .001$

[a] = team member ratings

Table 6 presents the correlations between the six measures of teamwork quality we extracted and the two measures of team performance, effectiveness and efficiency, rated by both team members and stakeholders. As expected, the correlations are all high and positive, with slightly higher correlations for the team members than for the stakeholders. These correlations were assessed with a structural equation model, following the procedures described in Hoegl & Gemuenden (2001), resulting in two models predicting evaluations of team performance by team members and stakeholders respectively.

In Model 1, we investigated the extent to which TWQ (evaluated by team members) predicts team members' evaluations of team performance (N = 199), whereas in Model 2 we investigated the extent to which TWQ (evaluated by team members) predicts stakeholders' evaluations of team performance (N = 199). The results of both models are presented in table 7. The results show also a good overall model fit for Model 1 (GFI = .997, AGFI = .993, RMR = .001, $\chi^2$ (19) = 1.417, $p$ = .99) and Model 2 (GFI = .997, AGFI = .995, RMR = .01, $\chi^2$ (19) = 0.873, $p$ = 1.0).

Table 7

*Results of the structural equation models at the individual level*

| | Model 1 TWQ Predicting Team Members' Evaluations of Team Performance | | Model 2 TWQ Predicting Stakeholders' Evaluations of Team Performance | |
|---|---|---|---|---|
| | Stand. Factor Loading | Stand. Coefficient/ *R*-Square | Stand. Factor Loading | Stand. Coefficient/ *R*-Square |
| Teamwork Quality (TWQ) | | | | |
|   Communication | .80 | | .79 | |
|   Coo. Of Exp. | .82 | | .82 | |
|   Trust | .74 | | .76 | |
|   Mutual Support | .84 | | .87 | |
|   Value Diversity | .84 | | .81 | |
| Team Performance | | 0.810.66 | | 0.64/0.40 |
|   Effectiveness | .79 | | .92 | |
|   Efficiency | .90 | | .83 | |
| GFI (Goodness-of-Fit Index) | | 0.997 | | 0.997 |
| AGFI (Ajusted GFI) | | 0.993 | | 0.995 |
| RMR (Root Mean Square Residual) | | 0.014 | | 0.011 |
| Chi-Square | | 1.417 | | 0.873 |
| Degrees of Freedom | | 19 | | 19 |
| *P*-Value | | .99 | | 1.0 |

*Note.* N = 199

While both models show a good fit, the relationship between TWQ and team performance is stronger for team members than for stakeholders. The same holds for the predictive power of the model (66% versus 40% explained variance). It makes intuitive sense that the relationship between team members own assessment of their team work and performance is stronger than that between their assessment of team work and their performance as assessed by stakeholders.

As mentioned, identical patterns emerge in the models at the team level (N = 29). However, the explained variance is substantially larger when using team ratings to predicts team performance as assessed by team members themselves (81% explained variance) or stakeholders (64% explained variance), possibly due to decreased variability at the team level.

To conclude, TWQ is significantly related to team performance measures of both team members and stakeholders. For perfomance as rated by team members, TWQ has a higher explanatory power than for performance rated by stakeholders.

## 5. DISCUSSION

### 5.1 DISCUSSION

The purpose of this study was to examine the factors within software development teams that have a significant influence on their performance to accomplish tasks. Even though a lot of time and money is being invested in software development, the success rate of projects is still disappointing (Wateridge, 1995; The Standish Group, 2009). Therefore it is important to know more about the factors that influence team performance. The focus of our study was on the quality of interactions within software development teams. The goal of the study was to test if trust, value sharing, and coordination of expertise contribute to explaining project success and to extend the model by Hoegl and Gemuenden (2001).

The study aimed to answer the research questions about (1) how teamwork quality is related to the success of software development projects and (2) how the new TWQ model compares to the original model of Hoegl and Gemuenden (2001). Our three main findings are: (1) There is a significant relationship between TWQ and the success of software development projects as measured by team performance (effectiveness and efficiency); (2) The magnitude of the relationship between TWQ and team performance differs with the perspective of the rater (team member versus stakeholder). TWQ explains 66% of the variance of team performance as rated by team members and 40% as rated by stakeholders; (3) The new TWQ model (which includes trust, coordination of expertise, and value sharing) explains team performance better (in terms of explained variance) than the TWQ model of Hoegl and Gemuenden (2001).

Results support the conceptualization of TWQ as latent construct, as the six factors loaded high on one factor in the factor analyses. This implies that the TWQ construct is a good measure of the collaborative nature of teams, focusing on the quality of the interactions within teams. The explained variance of the latent construct (TWQ) was almost eight percent higher than Hoegl and Gemuenden (2001) found for their model. This suggests that our adjustments have improved the model, implying that our model encompasses TWQ better than Hoegl and Gemuenden (2001).

Furthermore, the results of this empirical study show a significant relationship between TWQ and the success of software development projects. Based on data from 29 software development teams we found support for the relationship between TWQ and performance that was reported by Hoegl and Gemuenden (2001) and in other earlier studies regarding the relationship between individual aspects of TWQ and performance (Pinto et al., 1993; Lian et al., 2011; Faraj & Sproull, 2000). TWQ was found to relate significantly to team performance measures (effectiveness and efficiency). Therefore, the TWQ model can be used as a tool to evaluate the quality of interactions within teams and get useful information for team management. Based on the TWQ model, managers can adjust their management activities to improve team collaboration.

The size of the relationships between each TWQ factor and the team performance measures were found to be equal. No significant differences were found between the correlations of the six TWQ factors with the measures of effectiveness and efficiency for both team members and stakeholders (see Table 6). However, communication, mutual support and coordination of expertise showed somewhat higher correlations with performance measures. This suggests that communication and mutual support are of somewhat more importance for team performance than the other TWQ factors.

In general, TWQ explains 66% of the variance in team performance based on team member ratings and 40% of stakeholder ratings (as mentioned, the explanatory power of TWQ was found to be higher when the model was tested on the team level (81% and 64% respectively). However, in most cases, TWQ explains more than half of the variance of team performance. Hoegl and Gemuenden (2001) found a similar discrepance between the explanatory power of TWQ on team performance between team members and stakeholders.

There are several explanations for the difference in explanatory power between our TWQ model and that of Hoegl and Gemuenden (2001). First, our study was conducted 12 years after that of Hoegl and Gemuenden (2001). Many developments have taken place since. For example, after the "agile manifesto" was written in 2001 by Fowler and Highsmith, agile development methods have gained popularity and have been adopted by the industry more and more every year (Hussain, Slany, & Holzinger, 2009). Such changes in the industry make it difficult to compare our sample to that of Hoegl and Gemuenden (2001). Also, we do not make a distinction between team members, leaders and managers. We made the separation between individuals that are closely related with the day-to-day practices of the team (team members) and individuals that do have interest in the success of the project but are not part of the day-to-day practices (stakeholders). Finally, the research method was different. Hoegl and Gemuenden (2001) conducted on site interviews. Respondents were asked to complete the questionnaire while the interviewer was present to clarify any questions if needed. We used an online questionnaire that participants filled out in their own time, at their own place, without knowing the researchers. Both methods have their advantages and disadvantages, and might influence the results in their own way. Our high correlations may be related to the online questionnaire. The important note is that it is harder to compare results when research methods are different. Fourth, mean values of all the variables are close to ceiling in Hoegl and Gemuenden's (2001) study. Since we have no access to their data, we cannot know for sure, but it could be that the relatively high values affect the strength of their model.

To summarize, while we cannot directly compare the results of our study one-to-one to those of Hoegl and Gemuenden (2001), the results suggest that a model including trust, value sharing, and coordination of expertise contributes more to explaining project success than the factors of Hoegl and Gemuenden (2001) alone. Further research seems essential to be able to make strong(er) statements about this. Nevertheless, the significant relationship between TWQ and performance indicates the importance of managing team collaboration in software development teams. This knowledge is useful for software managers to build and manage teams more constructively and enhance performance.

Organizations spend large amounts of time and money on software development project. However, the success rate of these projects remains low. Much research is being done to investigate how the success rate of software development projects can be improved. Since software development primarily is a team effort and software quality has shown to be dependent on good teamwork, it is important to understand the characteristics of software development teams that significantly influence team performance.

We found evidence that better teamwork creates better team performance. Results of our study show that, indeed, there is a strong and significant relationship between teamwork quality and team performance in software development. Since our study is cross-sectional rather than longitudinal, we cannot verify causality of this relationship. However, based on the magnitude of the relationship, software managers cannot deny that teamwork quality is an important factor in achieving good performance. The TWQ model offers managers a way to assess teamwork quality and be able to reflect on it. Software managers should recognize the importance of teamwork quality and focus their management activities on actively improving the six TWQ factors.

This study has a number of limitations. First, the data for this research are from 28 Dutch and one Ukrainian software development teams. The regional scope of the study, therefore, is somewhat limited. However, the team level results support the individual level results, making it more likely that results can be generalized to the population. Nevertheless, generalization of the results only applies to the domain of software development. Software development teams need to work on complex tasks, in a constantly changing environment (Murray, 2000; Barry, Mukhopadhyay, & Slaughter, 2002). Good team collaboration is needed to be effective and work efficiently. When team tasks are less complex, teamwork quality might be a less important determinant for performance while other factors are. The TWQ model could be applicable in other domains, for example sports.

Second, being a cross-sectional study, results cannot provide definite information about the causality of the relationship between TWQ and performance. A longitudinal study could expand our knowledge about the causality of this relationship and about the development of team collaboration and the perception of performance over time.

Finally, as this article provides empirical evidence for the influence of TWQ on software team performance, further research should investigate the antecedents of TWQ. What can software managers do to enhance high team collaboration?

## CONCLUSION

Our goal in this study was to find additional factors that may influence software team performance, to extend the previous empirical TWQ study by Hoegl and Gemuenden (2001). We introduced three new TWQ factors: trust, value sharing, and coordination of expertise. The relationship between TWQ and team performance was tested using data from 252 team members and stakeholders. Results showed that teamwork quality is significantly related to team performance, as rated by both team members and stakeholders: TWQ explains 66% of the variance of team performance as rated by team members and 40% as rated by stakeholders. This study shows that trust, shared values, and coordination of expertise are important factors for team leaders to consider in order to achieve high quality software team work.

## REFERENCES

Agarwal, N. & Rathod, U. (2006). Defining 'success' for software projects: An exploratory revelation. *International Journal of Project Management, 24*(4), 358-370.

Baker, D. P., S. Gustafson, J. M. Beaubien, E. Salas & Barach, P. (2003). *Medical teamwork and patient safety: The evidence-based relation.* Washington, DC: American Institutes for Research.

Bandow, D. (2001). Time to create sound teamwork. *Journal for Quality and Participation*, 24, 41-47.

Barry, E. J., Mukhopadhyay, T. & Slaughter, S. A. (2002). Software project duration and effort: An empirical study. *Information Technology and Management, 3*(1-2), 113-136.

Bollen, K. A., & Hoyle, R. H. (1990). Perceived Cohesion : A Conceptual and Empirical Exandnation. *Social Forces*, *69*(2), 479–504.

Boss, R. W. (1978). Trust and managerial problem solving revisited. *Group Organization Management, 3*(3), 331–342.

Brodbeck, F. C. (2001). Communication and performance in software development projects. *European Journal of Work and Organizational Psychology, 10*(1), 73-94.

Campion, M. A., Medsker, G.J. & Higgs, A. C. (1993). Relations between work group characteristics and effectiveness: Implications for designing effective work groups. *Personnel Psychology, 46*(4), 823-850.

Cannon-Bowers, J. A., S. I. Tannenbaum, E. Salas & Volpe, C. E. (1995). Defining competencies and establishing team training requirements. In R. A.Guzzo & E. Salas et al. (eds*.). Team effectiveness and decision making in organizations,* pp. 333-380. San Francisco: Jossey-Bass.

Carron, A. V., Widmeyer, W. N. & Brawley, L. R. (1985). The development of an instrument to assess cohesion in sport teams: The group environment questionnaire. *Journal of Sport Psychology, 7*(3), 244-266.

Chang, A. & Bordia, P. (2001). A Multidimensional Approach to the Group Cohesion-Group Performance Relationship. *Small Group Research, 32*(4), 379-405.

Chin, W. W., Salisbury, W. D., Pearson, A. W. & Stollak, M. J. (1999). Perceived cohesion in small groups: Adapting and testing the perceived cohesion scale in a small-group setting. *Small Group Research, 30*(6), 751-766.

Chow, T. & Cao, D. B. (2008). A survey study of critical success factors in agile software projects. *Journal of Systems and Software, 81*(6), 961-971.

Cockburn, A. & Highsmith, J. (2001). Agile software development: The people factor. *Software Management, 34*(11), 131-133.

Cohen, S. G., Ledford, G. E. & Spreitzer, G. M. (1996). A predictive model of self-managing work team effectiveness. *Human relations, 49*(5), 415-437.

Cooke, R.A. & Szumal, J.L. (1994). The impact of group interaction styles on problem-solving effectiveness. *Journal of Applied Behavioral Science, 30*(4), 415-437.

Cooper, R. & Sawaf, A. (1996). *Executive EQ: Emotional intelligence in leadership and organizations.* New York: Grosset/Putnam.

Cooper, R. G. (1993). *Winning at new products: Accelerating the process from idea to launch.* Cambridge, MA: Addison Wesley.

Cooper, R. G. & Kleinschmidt, E. J. (1995). Benchmarking for firm's critical success factors in new product development. *Journal of Product Innovation Management, 39*(4), 1005-1023.

Creed, W.E.D. & Miles, R.E. (1996). Trust in organizations: A conceptual framework linking organizational forms, managerial philosophies, and the opportunity costs of controls. In R.M. Kramer & T.R. Tyler (eds.). *Trust in organizations: Frontiers of theory and research*, pp. 16-39. Thousand Oaks, CA: Sage.

Crowston, K., & Kammerer, E.E. (1998). Coordination and collective mind in software requirements development. *IBM Systems Journal, 37*(2), 227–245.

Delone, W.H. & McLean, E.R. (1992). Information Systems success: The quest for the dependent variable. *Information Systems Research, 3*(1), 60-95.

Delone, W.H. & McLean, E.R. (2003). The Delone and McLean Model of Information system success: A ten-year update. *Journal of Management Information Systems, 19*(4), 9-30.

Dodge, H.R., Fullerton, S. & Robbins, J.E. (1994). Stage of the organizational life-cycle and competition as mediators of problem perception for small businesses. S*trategic Management Journal, 15*(2), 121–134.

Dos Santos, B.L. (1986). A management approach to systems development projects. *Journal of Systems Management, 37*(8), 35–41.

Evans, J.R. & Mathur, A. (2005). The value of online surveys. *Internet Research, 15*(2), 195-219.

Faraj, S., & Sproull, L. (2000). Coordinating expertise in software development teams. *Management Science, 46*(12), 1554–1568.

Field, A. (2009). *Discovering statistics using SPSS.* London: Sage Publications.

Fowler, M. & Highsmith, J. (2001). The agile manifesto, http://www.agilemanifesto.org/

Friedlander, F. (1970). The primacy of trust as a facilitator of further group accomplishment. *Journal of Applied Behavioral Science, 6*(4), 387-400.

Furey, S. (1997). Why we should use function points. *American Menegement Systems, 14*(2), 28-30.

Gemuenden, H. G. (1990). Erflogsfaktoren des Projektmanagements: eine kritische Bestandsaufnahme der empirischen Untersuchungen. *Projekt Management, 90*(1,2), 4-15.

Gladstein, D. L. (1984). Groups in context: A model of task group effectiveness. *Administrative Science Quarterly, 29*(4), 499-517.

Griffin, A. & Hauser, J. R. (1992). Patterns of communication among marketing, engineering and manufacturing: A comparison between two new product development teams. *Management Science, 38*(3), 360-373.

Gully, S. M., Devine, D. J. & Whitney, D. J. (1995). A meta-analysis of cohesion and performance: Effects of level of analysis and task interdependence. *Small Group Research, 26*(4), 497-520.

Gupta, A. K., Raj, G. P. & Wileman, D. (1987). Managing the R&D marketing interface. *Resource Management,* (March), 38-43.

Hackman, J. R. (1987). The design of work teams. In J. Lorsch (eds.). *Handbook of organizational behavior*, pp. 315-342. New York: Prentice Hall.

Hackman, J.R. (1990). *Groups that work (and those that don't)*. San Francisco: Jossey-Bass.

Han, H. S., Lee, J. N. & Seo, Y. W. (2008). Analyzing the impact of a firm's capability on outsourcing success: A process perspective. *Information & Management, 45*(1), 31-42.

Handy, C. (1995). Trust and the virtual organization. *Harvard Business Review, 73*(3), 40–50.

Hauschildt, J. (1997). *Innovations management*, Muenchen: Franz Vahlen.

He, J., Butler, B.S. & King, W.R. (2007). Team cognition: Development and evolution in software project teams. *Journal of Management Information Systems, 24*(2), 261-292.

Henderson, J.C. & Lee, S. (1992). Managing I/S design teams: a control theories perspective. *Management Science, 38*(6), 757-777.

Hinsz, V.B., Tindale, R.S. & Vollrath, D.A. (1997). The emergent conceptualization of groups as information processors. *Psychology Bulletin, 121*, 43-64.

Hise, R. T., O'Neal, L., Parasuraman, A. & McNeal, J. U. (1990). Marketing/R&D interaction in new product development: Implications for new product success. *Journal of Product Innovation Management, 7*(2), 142-155.

Hoegl, M. & Gemuenden, H. G. (2001). Teamwork quality and the success of innovative projects: A theoretical concept and empirical evidence. *Organization Science, 12*(4), 435-449.

Hooper, D., Coughlan, J., & Mullen, M.R. (2008). Structural equation modelling: Guidelines for determining model fit. *Journal of Business Research Methods, 6*(1), 53–60.

Huang, S. J., & Han, W. M. (2008). Exploring the relationship between software project duration and risk exposure: A cluster analysis. *Information & Management, 45*(3), 175–182.

Hussain, Z., Slany, W. & Holzinger, A. (2009). Current state of agile user-centered design: A survey. *Lecture Notes in Computer Science, 5889*, 416 – 427.

Jarvenpaa & Leidner, 2006 Jarvenpaa, S. L., & Leidner, D. E. (2006). Communication and Trust in Global Virtual Teams. *Journal of Computer-Mediated Communication, 3*(4).

Jeffcott S. A. & Mackenzie, C. F.(2008). Measuring team performance in healthcare: Review of research and implications for patient safety. *Journal of Critical Care, 23*(2), 188-196.

Jehn, K. A. (1994). Enhancing effectiveness: an investigation of advantages and disadvantages of value-based intra-group conflict. *International Journal of Conflict Management, 5*(3), 223-238.

Jehn, K. A., Northcraft, G. B. & Neale, M. A. (1999). Why differences make a difference: A field study of diversity, conflict, and performance in workgroups. *Administrative Science Quarterly, 44*(4), 741-763.

Jones, G. & George, J. (1988). The experience and evolution of trust: Implications for cooperation and teamwork. *Academy of Management Review, 23*(3), 531-546.

Jones, M. C., & Harrison, A. W. (1996). IS project team performance: An empirical assessment. *Information & Management, 31*(2), 57–65.

Kaiser, H.F. (1974), An index of factorial simplicity, Psychometrika, 39(1), 31-36

Katz, R., & Allen, T. J. (1988). Investigating the not invented here (NIH) syndrome: A look at the performance, tenure, and communication patterns of 50 R&D project groups. In M.L. Tushman, W.L. Moore (eds.). *Readings in the Management of Innovations,* pp. 293-309. Cambridge, MA: Ballinger Publishing Company.

Kerrin, M. & Oliver, N. (2002). Collective and individual improvement activities: the role of reward systems. *Personnel Review, 31*(3), 320-337.

Kozlowski, S. W. J. & Klein, K. J. (2000). A multilevel approach to theory and research in organizations: Contextual, temporal, and emergent processes. In K. J. Klein & S. W. J. Kozlowski (eds.). *Multilevel theory, research, and methods in organizations*: Foundations, extensions, and new directions, pp. 3-90. San Francisco: Jossey-Bass.

Lei, P. W. & Wu, Q. (2007). Introduction to Structural Equation Modeling: Issues and Practical Considerations. *Educational Measurement: Issues and Practice*, 26(3), 33-43.

Lewis, K. (2003). Measuring transactive memory systems in the field: Scale development and validation. *Journal of Applied Psychology, 88*(4), 587–604.

Liang, T. P., Wu, J. C. H., Jiang, J. J., & Klein, G. (2012). The impact of value diversity on information system development projects. *International Journal of Project Management, 30*(6), 731-739.

Liu, J. Y. C., Chen, H. G., Chen, C. C. & Shue, T. S. (2011). Relationships among interpersonal conflict, requirements uncertainty and software project performance. *International Journal of Project Management, 29*(5), 547-556.

Lu, Y., Xiang, C., Wang, B. & Xiaopeng, W. (2010). What affects information systems development team performance? An exploratory study from the perspective of combined socio-technical theory and coordination theory. *Computers in Human Behavior, 27*(2), 811-822.

Mayer, R.C., Davis, J.H. & Schoorman, F.D. (1995). An integrative model of organizational trust. *Academy of Management Review, 20*(3), 709-734.

McGrath, J. E., Berdahl, J. L. & Arrow, H. (1996). No one has it but all groups do: Diversity as a collective, complex, and dynamic property of groups. In Susan E. Jackson and MarianN. Ruderman (eds.). *Diversity in Work Teams: Research Paradigms for a Changing World,* pp. 42-66. Washington, DC: APAPublications.

McGrath. J. E. (1964). *Social psychology: A brief introduction.* New York: Holt, Rinehart and Winston.

Morgan, B. B., Glickman, A. S., Woodward, E. A., Blaiwes, A. S. & Salas, E. (1986). Measurement of Team Behaviors in a Navy Environment (Technical Report No. NTSC TR-86-014). Orlando, FL: Naval Training Systems Center, Human Factors Division.

Mullen, B. & Copper, C. (1994). The relation between group cohesiveness and performance: An integration. *Psychology Bulletin, 115*(2), 210–227.

Murnighan, J. K., & Conlon, D. E. (1991). The dynamics of intense work groups: A study of British string quartets. *Administrative Science Quarterly, 36*(2), 165–186.

Murray, J. (2000). Reducing IT project complexity. *Information Strategy, 16*(3), 30–38.

Nonaka, I. & Takeuchi, H. (1995). *The Knowledge-Creating Company.* Oxford, UK: Oxford University Press.

O'Connor, M. A. (1993). The human capital era: Reconceptualizing corporate law facilitate labor-management cooperation. *Cornell Law Review, 78*(5), 899-965.

Pearce, J. L., Sommer, S. M., Morris, A., & Frideger, M. (1992). *A configurational approach to interpersonal relations: Profiles of workplace social relations and task interdependence.* Irvine: Graduate School of Management, University of California.

Pelled, L. (1996). Relational demography and perceptions of group conflict and performance: A field investigation. *International Journal of Conflict Resolution, 7*(3), 230-246.

Petter, S., DeLone, W.H. & McLean, E.R. (2008). Measuring Information systems success: Models, dimensions, measures, and Interrelationships. *European Journal of Information Systems, 17*, 236-263.

Pinto, M. B. & Pinto, J. K. (1990). Project team communication and cross-functional cooperation in new program development. *Journal of Product Innovation Management, 7*(3), 200-212.

Pinto, M. B., Pinto, J. K. & Prescott, J. E. (1993). Antecedents and consequences of project team cross-functional cooperation. *Management Science, 39*(10), 1281-1297.

Roberts, K., C. O'Reilly. (1974). Failures in upward communication in organizations: Three possible culprits. *Academy of Management Journal, 17*(2), 205–215.

Salas, E., C. A. Bowers & Cannon-Bowers, J. A. (1995). Military team research: 10 years of progress. *Military Psychology, 7*(2), 55–75.

Salas, E., Cannon-Bowers, J. A. & Johnston, J. H. (1997). How can you turn a team of experts into an expert team?: Emerging training strategies. In C. Zsambok & G. Klein (eds.). *Naturalistic decision making*, pp. 359-370. Hillsdale, NJ: Lawrence Erlbaum.

Salas, E., Cooke, N. J., and Rosen. (2008). On Teams, Teamwork, and Team Performance: Discoveries and Developments. *Human Factors: Journal of Human Factor Ergonomics Society, 50*(3), 540-547.

Salas, E., Sims, D. E., and Burke, C. S. (2005). Is there a "Big Five" in teamwork? *Small Group Research, 36*(5), 555-599.

Salas, E., Stagle, K. C., Burke, C. S., & Goodwin, G. F. (2007). Fostering team effectiveness in organizations: Toward an integrative theoretical framework of team performance. In R. A. Dienstbier, J. W. Shuart, W. Spaulding, & J. Poland (eds.). *Modeling complex systems: Motivation, cognition and social processes:* Nebraska Symposium on Motivation, Lincoln: University of Nebraska Press.

Sauer, C., Gemino, A., & Reich, B. H. (2007). The impact of size and volatility on IT project performance. *Communications of the ACM, 50*(11), 79–84.

Siau, K., Tan, X. & Sheng, H. (2010). Important characteristics of software development team members: an empirical investigation using Repertory Grid. *Information Systems Journal, 20*(6), 563-580.

Thamhain, H. J. & Kamm, J. B. (1993). Top-level managers and innovative R&D performance. A. Cozijnsen, W. Vrakking (eds.). *Handbook of Innovation Management*, pp. 42-53. Oxford, U.K.: Blackwell Publishers.

The Standish Group (2009). Chaos Summary Report 2009: The 10 laws of CHAOS. Technical Report. Boston, MA: The Standish Group International.

Tjosvold, D. (1995). Cooperation theory, constructive controversy, and effectiveness: Learning from crisis. R.A. Guzzo, E. Salas and Associates (eds.), *Team Effectiveness and Decision Making in Organizations,* pp. 79-112. San Fransisco, CA: Jossey-Bass.

Wang, E. T. G., Chang, J. Y. T., Jiang, J. J. &Klein, G. (2011). User advocacy and information system project performance. *International Journal of Project Management, 29*(2), 146-154.

Ward, E. A. (1997). Autonomous work groups: A field study of correlates of satisfaction. *Psychology Reports, 80,* 60–62.

Wateridge, J. (1995). IT projects: A basis for success. *International Journal of Project Management, 13*(3), 169-172.

Weick, K., & Roberts, K. (1993). Collective mind in organizations: Heedful interrelating on flight decks. *Administrative Science Quarterly, 38*(3), 357–381.

Zand, D. (1972). Trust and managerial problem solving. *Administrative Science Quarterly, 17*(2), 229–239.

**Appendices**

**Appendix A: Overview of Team Characteristics**

Table 8 *Overview of all team characteristics*

| Org. | Team | n | Team size | Project duration | Main assignment | Type of application | Application size | Development method |
|------|------|---|-----------|------------------|-----------------|---------------------|------------------|--------------------|
| 1 | 1 | 3 | 5-10 | 6-12 | Renovation | **Other*** | 10-100 | Incremental / **Agile**** |
| 2 | 2 | 6 | 5-10 | > 18 | Maintenance | Business | - | Agile |
| 3 | 3 | 16 | > 10 | < 6 | Maintenance | Business | - | Waterfall |
|  | 4 | 4 | > 10 | 6-12 | Maintenance | Real-time | - | **Waterfall**** / RUP |
|  | 5 | 18 | > 10 | < 6 | **New development*** | Business | - | Waterfall |
| 4 | 6 | 11 | > 10 | > 18 | New development | Business | 1.000-10.000 | Agile |
| 5 | 7 | 4 | **5-10** | < 6 | New development | Business | 100-1.000 | Agile |
| 6 | 8 | 12 | 5-10 | < 6 | Maintenance | Business | - | **Waterfall**** / Agile |
| 7 | 9 | 3 | < 5 | < 6 | New development | Other | - | **Incremental**** / Agile |
| 8 | 10 | 5 | 5-10 | > 18 | New development | Business | - | Agile |
| 9 | 11 | 8 | 5-10 | > 18 | New development | Business | 100-1.000 | RUP / **Agile**** |
|  | 12 | 4 | < 5 | > 18 | **Renovation**** / Maintenance | Business | 1.000-10.000 | Incremental |
|  | 13 | 6 | 5-10 | 6-12 | New development | Business | 100-1.000 | Agile |
|  | 14 | 25 | > 10 | > 18 | **Renovation*** | Business | 1.000-10.000 | Agile |
| 10 | 15 | 11 | 5-10 | 6-12 | New development | Business | > 10.000 | Agile |
| 11 | 16 | 4 | < 5 | < 6 | New development | Business | > 10.000 | Incremental |
| 12 | 17 | 3 | < 5 | > 18 | New development | Business | > 10.000 | **Incremental\*\*** / Agile |
| 13 | 18 | 5 | < 5 | 6-12 | New development | Real-time | 1.000-10.000 | Waterfall |
| 14 | 19 | 9 | 5-10 | > 18 | **New development\*\*** / Maintenance | Business | - | Agile |
| | 20 | 6 | 5-10 | **6-12** | New development | Business | > 10.000 | Agile |
| 15 | 21 | 4 | 5-10 | > 18 | System renovation | Business | 100-1.000 | Waterfall |
| | 22 | 6 | 5-10 | > 18 | Maintenance | Business | - | **Waterfall\*\*** / Incremental / Agile |
| | 23 | 2 | **5-10** | **6-12** | **Other\*** | **Other\*** | - | **Waterfall\*** |
| 16 | 24 | 3 | < 5 | < 6 | New development | Business | - | Waterfall / Incremental / **Agile\*\*** |
| 17 | 25 | 5 | 5-10 | 6-12 | New development | Business | - | Agile |
| | 26 | 2 | **< 5** | **< 6** | Re-development | **Business\*** | - | Agile |
| | 27 | 4 | > 10 | 6-12 | New development | Business | - | Waterfall |
| 18 | 28 | 3 | < 5 | < 6 | New development | Other | 100-1.000 | Agile |
| | 29 | 7 | 5-10 | > 18 | **Renovation\*** | **Other\*** | - | Agile |

\* Answer with highest response rate when no majority (> 50%) selected the same answer.

\*\* Multiple answers were selected by the majority (> 50%), this is the answer with the highest response rate.

**Appendix B: Graphical Representation of the TWQ Model**

Figure 4. *Graphical representation of the TWQ model in AMOS*

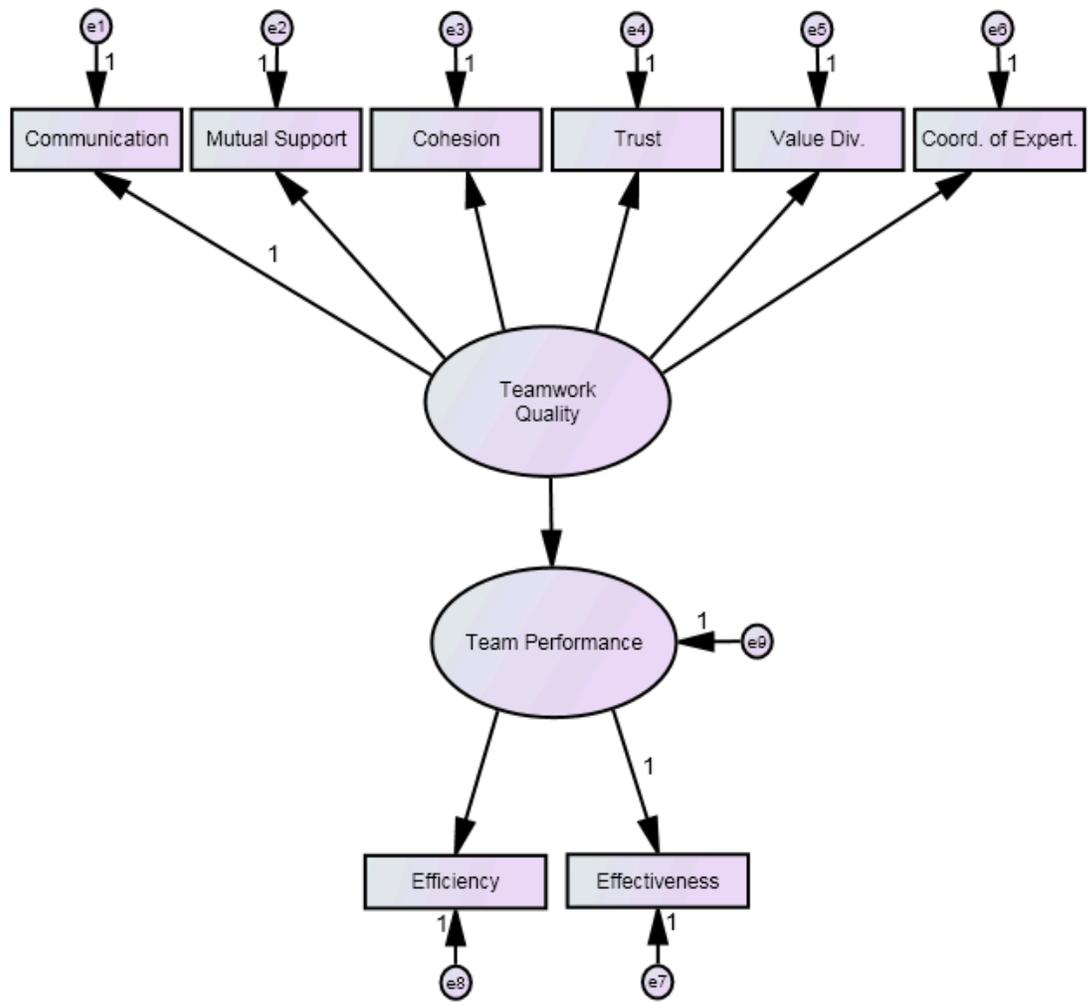